\documentclass[twocolumn,showpacs,preprintnumbers,amsmath,amssymb,prl]{revtex4}

\usepackage{graphicx}
\usepackage{dcolumn}
\usepackage{bm}

\begin{document}

\title{Liquid-liquid phase transition of water in hydrophobic and hydrophilic pores}

\author{I. Brovchenko}
\email{brov@pc2a.chemie.uni-dortmund.de} 
\author{A. Oleinikova}
\affiliation{Physical Chemistry, University of Dortmund, Otto-Hahn-Str.6, Dortmund, D-44227, Germany
}

\date{\today}
\begin{abstract}
Effect of confinement on the liquid-liquid transition of water are studied by simulations in the Gibbs ensemble. Upon cooling along the liquid-vapor coexistence curve, confined water undergoes transition from normal to strongly tetrahedral water via a first order phase transition (as in the bulk) or in a continuous way in dependence on pore hydrophilicity. In all cases, transition temperature is only slightly shifted by the confinement. This agrees with the experimentally observed weak effect of confinement on the temperature of the fragile-to-strong transition of water.    
\end{abstract}
\maketitle
The fragile-to-strong dynamic transition (FST) of bulk liquid water at about 228 K \cite{Angell99} originates from the change of water structure due to the thermodynamic liquid-liquid phase transition from normal water to strongly tetrahedral water \cite{Poole92}. Crystallization prevents direct experimental studies of this phase transition and location of its critical point remains unknown. In water models with a liquid-liquid critical point at positive pressure, a sharp change of water properties upon cooling at zero pressure corresponds to the crossing the line of the heat capacity maxima \textit{C$_p^{max}$} \cite{Poole05,Stanley2005}. In other models, the liquid-liquid critical point is located at negative pressures and water undergoes a first order phase transition upon cooling at zero pressure \cite{mult1,mult2}. Water crystallization can be suppressed by confinement and studies of the supercooled confined water may clarify the location of the liquid-liquid transition of bulk water. Recent experiments \cite{Zanotti,Chen2004,Chen2005,Mallamace,Chen2006a,Chen2006b} show that FST of water in various confinements occurs in relatively narrow temperature interval close to 228 K, where FST is expected for bulk water. In the incompletely filled Vycor pores, confined water undergoes first order liquid-liquid transition at 240 K, accompanied by the strong changes of water dynamics \cite{Zanotti}. In cylindrical silica mesopores, FST occurs at about 225 K at zero pressure and its shift with pressure follows the behavior expected for the liquid-liquid transition \cite{Chen2004,Chen2005,Mallamace}. In more hydrophobic carbon nanotubes, water shows FST at slightly lower temperatures (about 218 K) \cite{Mamontov}. Interestingly, the dynamic transition of various hydrated biomolecules, accompanied by the onset of their function upon heating, occurs approximately at the same temperatures (200 to 230 K) \cite{Sokolov}. Obviously, dynamic transition of biomolecules is governed by FST of hydration water, which occurs at 222 K and 220 K in the case of DNA \cite{Chen2006b} and lysozyme \cite{Chen2006a} molecules, respectively. 
\begin{figure}[b]
\includegraphics[width=7.5cm]{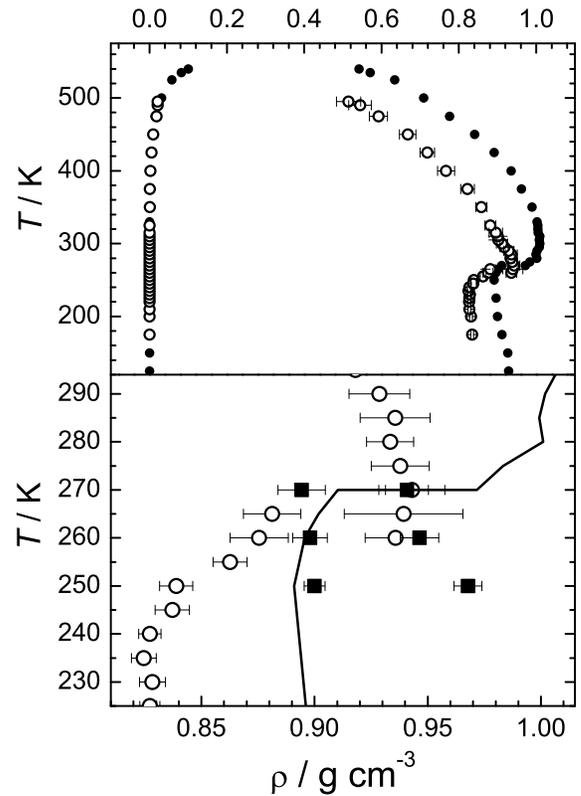}
\caption{Phase diagram of water in hydrophobic pore (\textit{U}$_0$ = -0.39 kcal/mol): liquid-vapor coexistence (open circles), liquid-liquid coexistence (squares). Liquid-vapor coexistence curve of the bulk water is shown by solid circles and solid line.}
\end{figure} 
\par
The phase transitions of fluids are strongly affected by confinement and the robustness of the FST temperature of water with respect to various confinements seems to be surprising. Density functional studies show that the liquid-liquid transition of water-like fluid in hydrophobic pore is weakly affected by confinement only when the fluid is assume to be homogeneous  \cite{Deb01}. Computer simulations of model water in very narrow pores indicate possibility of the liquid-liquid transitions of confined water \cite{Stanleypore,Kumar2005}, but liquid-liquid transition of confined model water was not found yet. To clarify the effect of confinement on the liquid-liquid phase transition of water, we have performed simulation studies of ST2 water \cite{ST2}, whose bulk phase diagram is known in details \cite{mult2}, in slit-like pores of 24 $\mbox{\AA}$ width. Contrary to the bulk case \cite{mult1,mult2}, we did not use long-range corrections for Lennard-Jones (LJ) intermolecular interactions in pore geometry. The effect of these corrections on the bulk phase diagram of ST2 water was found negligible at \textit{T} $\geq$ 250 K, whereas it causes increase of liquid water density on about 1.5 $\%$ at lower temperatures. Interaction of smooth pore wall with water oxygens was represented by (9-3) LJ potential, whose well-depth \textit{U}$_0$ was varied to reproduce hydrophobic (\textit{U}$_0$ = -0.39 kcal/mol),  moderately hydrophilic (\textit{U}$_0$ = -1.93 kcal/mol) and strongly hydrophilic (\textit{U}$_0$ = -3.08 kcal/mol) pore walls. 
\begin{figure}[htb]
\includegraphics[width=7.5cm]{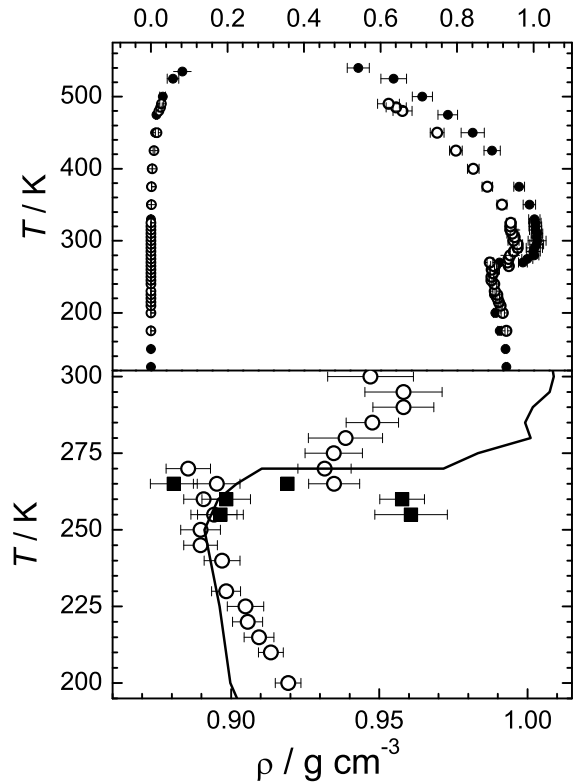}
\caption{Phase diagram of water in moderately hydrophilic pore (\textit{U}$_0$ = -1.93 kcal/mol): liquid-vapor coexistence (open circles), liquid-liquid coexistence (squares). Liquid-vapor coexistence curve of the bulk water is shown by solid circles and solid line. }
\end{figure}
\par
The liquid-vapor coexistence of water was simulated by direct equilibration of the two coexisting phases in the Gibbs ensemble \cite{GE}. The liquid-liquid phase transition was detected by the appearance of a triple point, where vapor coexists with two liquid phases of different densities. At the triple point, the liquid branch of the liquid-vapor coexistence curve shows a pronounced step in density \cite{mult1,mult2}. Additionally, coexistence of two liquid phases close to the triple point was studied by direct equilibration in the Gibbs ensemble. The use of efficient techniques for the molecular transfers (more details can be found elsewhere \cite{water1,mult2}) allowed extension of the simulated liquid-vapor coexistence curves deeply into the supercooled region and direct equilibration of two liquid phases at \textit{T} $\ge$ 250 K. The total number \textit{N} of molecules in the two simulation boxes used for liquid-vapor and liquid-liquid equilibration was about 850 and 1600, respectively. The number of successful molecular transfers in the course of the simulation runs always exceeded 10 \textit{N} and achieved 100 \textit{N} at high temperatures for liquid-vapor equilibration, whereas for liquid-liuqid equilibration it was about 2 \textit{N}.
\par
The liquid-vapor coexistence curves of ST2 water in hydrophobic and moderately hydrophilic pores are shown in Figs.1 and 2, respectively. In both cases the liquid density shows step-like density change by 0.05 to 0.06 g cm$^{-3}$ at \textit{T} $\approx$ 260 - 270 K. This density step indicates a triple point, where a saturated vapor coexists with two liquid phases of different densities. A hysteresis observed in the temperature range of about 10$^\circ$ clearly evidences the first order liquid-liquid phase transition. This finding was confirmed by the direct equilibration of the two coexisting liquid phases in the Gibbs ensemble at \textit{T} $\geq$ 250 K (see squares in Figs.1 and 2). Similarly to the bulk case \cite{mult1,mult2}, the liquid-liquid phase transition of confined water shifts to lower pressures (densities) with increasing temperature and should end at the critical point at some negative pressure. The temperatures \textit{T}$_{t}$ of the liquid-liquid-vapor triple point estimated as a middle of the hysteresis loop are shown in Table I. Evidently, the temperature of the triple point decreases by just a few degrees due to the confinement in hydrophobic and moderately hydrophilic pores. Note, that in both pores \textit{T}$_{t}$ is essentially above the glass transition temperature \textit{T}$_g$, estimated from the temperature dependence of the heat capacity of liquid along the liquid-vapor coexistence curve (see Table I).
\begin{table}[b]
\caption{Temperature \textit{T}$_t$ of the liquid-liquid-vapor triple point and glass transition temperature \textit{T}$_g$ of bulk and confined ST2 water. The temperatures \textit{T}$_t^*$ = \textit{T}$_t$ - 30$^{\circ}$ are expected for real water, when the temperature shift of the liquid density maximum in ST2 water is taken into account.}
\label{tab:par} 
\vspace{0.5cm}     
\begin{tabular}{ccccc}
\hline\noalign{\smallskip}
System&\textit{U}$_0$ \ \ & \ \ \ \ \ \ \textit{T}$_t$ \ \ \ \  & \ \ \ \ \textit{T}$_t^*$ \ \ \ &\ \ \ \textit{T}$_g$ \ \ \ \\
 &(kcal/mol)&(K)&(K)&(K)\\
 \noalign{\smallskip}\hline\noalign{\smallskip}
   ST2, bulk \ \ \ \ \ \ \ & - &$\sim$ 270&$\sim$ 240 &$\sim$ 235 \\
 \ \ \textit{H} = 24 $\mbox{\AA}$ &-0.39 & $\sim$ 263 & $\sim$ 233 &$\sim$ 225 \\
  \ \ \textit{H} = 24 $\mbox{\AA}$ &-1.93 & $\sim$  268 & $\sim$ 238 &$<$ 200 \\
  \ \ \textit{H} = 24 $\mbox{\AA}$ &-3.08 & $\sim$ 275\footnotemark[1] & $\sim$ 245\footnotemark[1] &$<$ 200  \\
\noalign{\smallskip}\hline\noalign{\smallskip}
 ST2RF, bulk  \ \ \ \ \ \ & - & $\sim$ 280\footnotemark[1]&& $\sim$ 255 \\
\noalign{\smallskip}\hline
\end{tabular}
\footnotetext[1]{heat capacity maxima \textit{C$_p^{max}$}}
\end{table} 
\begin{figure}[htb]
\includegraphics[width=7.5cm]{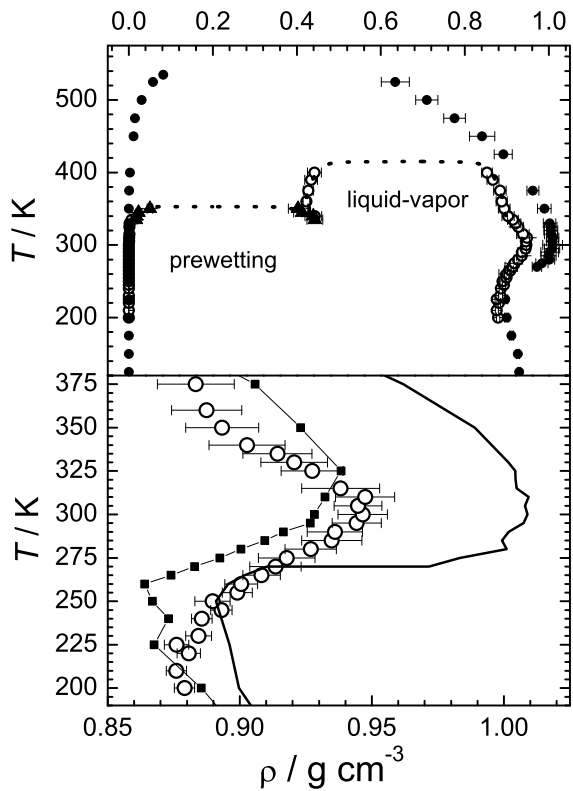}
\caption{Phase diagram of water in strongly hydrophilic pore (\textit{U}$_0$ = -3.08 kcal/mol): liquid-vapor coexistence of water in pore interior (open circles), prewetting transition (solid triangles). Dotted lines show approximately the shape of the coexistence curve expected for 2D transitions. Bulk liquid-vapor coexistence curves are shown for ST2 water (solid circles and solid line) and ST2RF water (squares). }
\end{figure}
\par
Further strengthening of the pore wall hydrophilicity causes appearance of a prewetting transition, which is a two-dimensional condensation of a thin water film (of about two water layers in the case of nanopores \cite{water1}). As a result, in strongly hydrophilic pore liquid-vapor phase transition splits into the prewetting transition and the liquid-vapor transition in a pore with a wall already covered by about 2 water layers (Fig. 3). The liquid density maximum is strongly enhanced in comparison with a bulk water and liquid density changes gradually without any indications on the existence of a liquid-liquid-vapor triple point (Fig.3, lower panel). Such behavior should be attributed to the shift of the liquid-liquid critical point to higher (positive) pressures. Similar shift of the liquid-liquid critical point was observed for the simple van-der-Waals model of hydrogen-bonded liquid \cite{vdW} with the strengthening of hydrogen bonds and also for ST2 water model with orientational ordering enhanced by the application of the long-range corrections for the intermolecular Coulombic interaction (ST2RF model \cite{mult2}, see lower panel in Fig.3). In the studied strongly hydrophilic pore, we may attribute the shift of the liquid-liquid critical point to the strong enhancement of the orientational ordering of water molecules near hydrophilic surfaces \cite{water1}. Liquid branch of the liquid-vapor coexistence curves of ST2RF bulk water and of ST2 water in strongly hydrophilic pore look like superctitical isobar of the liquid-liquid phase transition close to its critical point (Fig.3, lower panel). The inflection of this isobar indicates crossing the line of the heat capacity maxima \textit{C$_p^{max}$} emanating from the liquid-liquid critical point \cite{Poole05,Stanley2005}, which is located at about 280 K for bulk ST2RF water \cite{Poole05,mult2}. For ST2 water in strongly hydrophilic pore \textit{C$_p^{max}$} is located at about 275 K. So, transition between two liquid phases of confined water at zero pressure occurs very close to the temperature of the liquid-liquid-vapor triple point of bulk ST2 water (Table I). 
\par
Structural analysis of two liquid phases of bulk ST2 water, which may coexist with vapor, shows that four-coordinated tetrahedrally ordered water molecules dominates in the low-density phase (phase I), whereas normal density water (phase II) is enriched with molecules, which have tetrahedrally ordered four nearest neighbors and up to 6 molecules in the first coordination shell \cite{mult2006}. Near hydrophobic surface, these two phases show quite different arrangement even in the surface layer (Fig.4). Enhanced tetrahedral ordering in the phase I causes splitting of the water surface layer near hydrophobic surface (Fig.4,a). Near the moderately hydrophilic wall, the surface layer remains strongly localized and its density only slightly decreases upon liquid-liquid transition (Fig.4,b). In strongly hydrophilic pore, transition to low-density water causes decrease of density in the pore center, whereas the surface layer remains unchanged in a wide temperature range (Fig.4,c). So, we may expect that the liquid-liquid transition of water near heterogeneous biological surfaces, containing both hydrophobic and hydrophilic areas, should affect first of all structural and dynamical properties of water near hydrophobic parts of biomolecule.  
\begin{figure}[b]
\includegraphics[width=7cm]{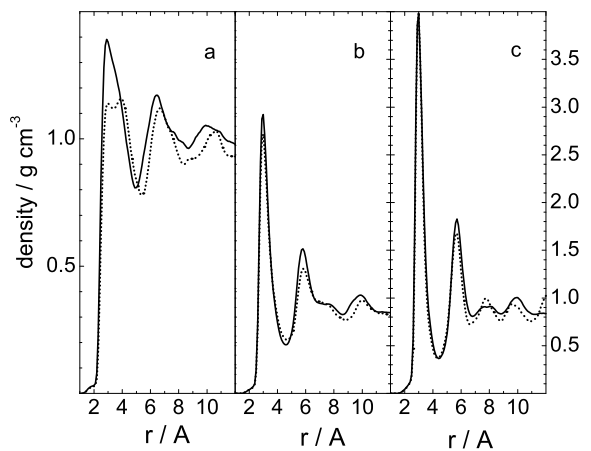}
\caption{Density profiles of two coexisting liquid water phases in hydrophobic (\textbf{a}, \textit{T} = 265 K, vertical scale is enhanced) and moderately hydrophilic (\textbf{b}, \textit{T} = 270 K) pores. Density profiles of liquid water phases in strongly hydrophilic pore (\textbf{c}, \textit{T} = 300 K: solid line, \textit{T} = 260 K: dotted line). }
\end{figure}  
\par
Confinement affects strongly the liquid-vapor phase transition of ST2 water and its critical temperature decreases by about 50$^\circ$ in strongly hydrophobic and moderately hydrophilic pores (Figs.1 and 2) and even more in strongly hydrophilic pores (Fig.3). In contrast, the temperature of the liquid-liquid-vapor triple point (or the temperature of \textit{C$_p^{max}$}) is much less sensitive to the confinement. This may be directly related to the large slope of the liquid-liquid transition line in temperature-pressure plane with respect to zero pressure line. As a result, the pressure shift of the liquid-liquid coexistence in confinement causes only small shift of \textit{T}$_t$ or the temperature of \textit{C$_p^{max}$}. This shift is negative in hydrophobic and moderately hydrophilic pore and reflects stronger disordering effect of a surface on more ordered strongly tetrahedral water phase. The temperature shift is opposite in strongly hydrophilic confinement, which promotes more ordered low-temperature water phase due to the strong orientational ordering of water molecules near hydrophilic surfaces. The observed trends agree with higher temperature of FST of water in hydrophilic Vycor glass ($\sim$ 240 \cite{Zanotti}) and silica pores ($\sim$ 225 K \cite{Chen2004,Chen2005,Mallamace}) relatively to the temperature of FST in less hydrophilic carbon nanotubes ($\sim$ 218 K \cite{Mamontov}). 
\par
Our results indicate, that by tuning the pore hydrophilicity the liquid-liquid critical point may be placed exactly at the liquid branch of the liquid-vapor coexistence curve. Experimental studies of supercooled water in pores of various hydrophilicity may clarify, whether the temperature of real bulk water singularities (about 228 K) corresponds to the liquid-liquid-vapor triple point or to the distant effect of the liquid-liquid critical point located at positive pressure. Note, that the liquid-liquid phase transition studied in the present paper is the lowest-density one among the multiple liquid-liquid transitions of water seen in experiment \cite{Loert06} and simulations \cite{mult1,mult2,Jedl}. In particular, with increasing pressure, ST2 water undergoes transition from normal density water to much less orientationally ordered liquid \cite{mult2006,perc2006}. Further studies should clarify effect of this high-pressure transition on the properties of confined water and its probable relation to the pressure-induced denaturation of biomolecules \cite{Heremans}.    
\par
We thank the Deutsche Forschungsgemeinschaft (DFG-Forschergruppe 436) for financial support. 

\end{document}